\documentclass[a4paper,12pt]{article}

\usepackage{amsmath}
\usepackage{amsfonts}
\usepackage{amssymb}
\usepackage{a4wide}
\def\c #1{{\cal #1}}                             
\def\Dirac{{\raise0.09em\hbox{/}}\kern-0.69em D}
\def\ep{i\epsilon} 

\def\kbar{{\mathchar'26\mkern-9muk}} 		
\def\lesssim{\mathrel{\hbox{\rlap{\hbox{\lower4pt\hbox{$\sim$}}}\hbox{$<$}}}}
\def\sq{\hbox{\rlap{$\sqcap$}$\sqcup$}}         
\def\p{\partial}                                
\def\tfrac #1#2{\textstyle{\frac{#1}{#2}}} 	
\def\tr{\mbox{Tr}\,}                            

\def\beg{\begin{eg}\rm}                         
\def\eeg{\hfill\sq\end{eg}}                     

\def\cF{{\cal F}}      
\def\cS{{\cal S}}
 \def\cG{{\cal G}} 
 
\def\k {\kern-.1em\mathbin{,}\kern-.1em}
\def\hk{\kern.12em\raise-1em\hbox{$\hat{\raise1em\hbox{,}}$}\kern.12em}
  
\def\cA{{\bf A}}

\def\exterior{{{\raise0.2em\hbox{$\scriptstyle\bigwedge$}}{}}}

\def\tA{{\sf A}}

\def\tF{{\sf F}}

\def\tX{{\sf X}}

\newcounter{eg}                                 
\newtheorem{eg}{Example}[section]
\def\beg{\begin{eg}\rm}                         
\def\eeg{\hfill\sq\end{eg}}                     
        
\newcommand{\initiate}{\setcounter{equation}{0}}        
\begin{document}

\title{Gauge fields on noncommutative geometries \\ 
        with curvature}

\vskip30pt
\author{
M.~Buri\' c$^{1}$\thanks{majab@phy.bg.ac.rs},\  \ 
H.~Grosse$^{2}$\thanks{harald.grosse@univie.ac.at}\ \ 
                    and
J.~Madore$^{3}$\thanks{madore@th.u-psud.fr}  \\[25pt]
\\$\strut^{1}$
      Faculty of Physics,
       University of Belgrade \\ Studentski trg 12,
       SR-11001 Belgrade 
 \\[10pt]$\strut^{2}$
Department of Physics,
University of Vienna
\\ Boltzmanngasse 5, A-190 Vienna
                   \\[10pt]$\strut^{3}$
       Laboratoire de Physique Th\'eorique,
       Universit\'e de Paris-Sud \\
B\^atiment 211, F-91405 Orsay}  
\maketitle

\begin{abstract}
It was shown recently that the lagrangian of the
Grosse-Wulkenhaar model can be written as lagrangian of the
scalar field propagating in a curved noncommutative space.
In this interpretation, renormalizability of the model is 
related to the interaction with the background curvature
which introduces explicit coordinate dependence in the
action. In this paper we construct the $U_1$ gauge field 
on the same noncommutative space: since covariant derivatives
contain coordinates, the Yang-Mills action is again 
coordinate dependent. To obtain a two-dimensional 
model we reduce to a subspace, which results in
splitting of the degrees of freedom into a gauge 
and a scalar. We define the gauge fixing  
and show the BRST invariance of the quantum action.
\end{abstract}

\initiate
\section{Introduction}

Gauge theories were first formulated on matrix geometries late in the
last century~\cite{gauge}. At first interest was concentrated on
their properties as classical field theories and mainly based on the
fact that the extension {\it \`a la} Kaluza-Klein of an ordinary
geometry by an algebra of $n \times n$ matrices transforms
electromagnetism, a $U_1$-theory, into a Yang-Mills-Higgs-Kibble
$(S)U_n$ theory.  Subsequent to the work of Seiberg and Witten at the
dawn of the new century~\cite{sw} the interest of the majority of
workers in the field focused on Moyal spaces~\cite{chep}, partly
because of the unique algorithm which gives the description of field
theories. It was soon realized that upon quantization the gauge fields
are plagued with the same $UV/IR$--mixing behavior as the scalar field,
that is, although the resulting models remain as they should $UV$--finite, 
the noncommutativity gives rise to a new $IR$~divergence.

This renormalizability problem has been recently solved in the case of
scalar field by Grosse and Wulkenhaar (GW) by including an
additional term of the form $x^2\phi^2$ in the action, \cite{gw}.
The oscillator term provides the symmetry between short and long
distances or between coordinates and momenta which is referred to as
Langmann-Szabo (LS)~duality, \cite{ls}. The external potential in the 
GW action breaks
translational invariance; scalar models which are translationally
invariant and LS~dual were subsequently proposed and proved to be
renormalizable, \cite{tran}. They generically contain an additional
term $\p^{-2}\phi^2$ in the action which changes the form of the
scalar field propagator to $(p^2 + m^2 +\frac{a}{p^2} )^{-1}$.

One would like to be able to construct renormalizable models for the
gauge fields in a similar way. However, there is a 
problem which has not been
satisfactorily solved: how to include the confining 
coordinate-dependent terms in covariant or systematic manner. 
There are  various proposals which
we recall  briefly here, for more detailed
reviews see \cite{Blaschke:2009rb,de Goursac:2007gq}.  
Possibility which the authors of \cite{de Goursac:2007gq,Grosse:2007dm} 
took is to start from a scalar field interacting with an external
gauge field, and then integrate the scalar degree of freedom. In this
manner they obtained the `induced gauge model': it is expressed in 
terms  of the so-called covariant coordinates and includes thus
spatial coordinates in a natural way.
However the model does not have a trivial vacuum so it is not
clear how to quantize, \cite{deGoursac:2008rb,Grosse:2007jy}.  
The other possibility to introduce
coordinate dependence  is through the ghost sector:
 such a model was described in \cite{Blaschke:2007vc}, where  also its
BRST invariance was proved.  The $\p^{-2}$ gauge theories were defined
and explored in considerable details in \cite{Blaschke:2009hp}.
However a full renormalizability analysis of mentioned models
for various reasons is still missing.

In recent paper \cite{trHei} we proposed a geometric interpretation 
of the oscillator term: It was shown there that the 
Grosse-Wulkenhaar action can be interpreted as an action 
for the scalar field coupled to the curvature of a background 
noncommutative space.  The coupling has the usual form, $R\phi^2$,
and in this term the oscillator potential is contained.
In this paper we explore further the 
differential structure of the mentioned background space 
(which we call the truncated Heisenberg algebra) and
we construct the $U_1$ gauge theory on it: the resulting
action presents geometric analog of the GW action for the gauge 
fields. We also initiate the study of renormalizability  
by proving the BRST invariance of the gauge fixed action.

The plan of the paper is the following: in Section~2 we
recollect some results of \cite{trHei} and also some steps 
of the construction of  local symmetries in the 
noncommutative frame formalism,
\cite{gauge,book}. We apply the formalism  to 
the truncated Heisenberg space and 
then we reduce to subspace  $z=0$ which gives the
relevant two-dimensional theory in Section~3.
 In Sections 
4 and 5 we discuss the Yang-Mills and the Chern-Simons 
actions and the corresponding classical equations of motion. 
Finally in the last section we specify the gauge fixing 
and we show that the quantum action is BRST invariant.
Note that, in notation which we use,
coordinates and their functions (fields) are generically operators 
and therefore the product is always noncommutative: depending on the
representation it is either matrix product or the Moyal product.
Similarly the trace denotes respectively the matrix trace or 
the integral.

\initiate
\section{Truncated Heisenberg algebra}

Generically a noncommutative space is an algebra $\c{A}$ generated by
a set of hermitian elements which we shall loosely refer to as
`coordinates'.  The truncated Heisenberg space, \cite{trHei}, is a 
three-dimensional noncommutative space defined by 
coordinates $x$, $y$ and $z$ and the commutation relations
\begin{eqnarray}
&&
[x,y] = i \epsilon \mu^{-2}(1-\mu z) ,  \nonumber\\[4pt]
&& [x,z]  = i\epsilon(yz+zy) ,          \label{alg}\\[4pt]
&& [y,z] = - i\epsilon(xz+zx) .        \nonumber
\end{eqnarray} 
The $\mu$ is a constant of dimension of the inverse length;
physically in fact it would make sense (and be
consistent) to introduce two different length
scales, $\mu$ and $\bar\mu$, in (\ref{alg}): $\mu$  as 
a characteristic dimension of  $x$-$y$ space, and $\bar\mu$
for the auxilliary $z$-direction. For simplicity however
we keep $\mu$ and $\bar\mu$ the same. The
$\epsilon $ is a dimensionless parameter which indicates the relative
strength of noncommutativity; we denote $\kbar = \epsilon \mu^{-2}$.
For $\epsilon =1$ algebra (\ref{alg}) can be represented by 
$n\times n$ matrices for any integer $n$,
\begin{equation}
x = \frac{1}{\mu \sqrt{2}}
\begin{pmatrix}
0 & 1 &0 &.& . &. \cr
1 &0 &\sqrt{2}& .&. &. \cr
0  &\sqrt{2} & 0&. &.& .\cr
. &. &.  &. & . &. \cr
. & . & .&. &0 & \sqrt{n-1} \cr
. & . & .& .&\sqrt{n-1}& 0 
\end{pmatrix} ,
\end{equation} 

\begin{equation}
 y = \frac{i}{\mu \sqrt{2}}
\begin{pmatrix}
0 & -1 &0 & . &.&. \cr
1 &0 &-\sqrt{2}& .&. &. \cr
0  &\sqrt{2} & 0&. &.&.\cr
. &. &.  &. & . &. \cr
. & . & . &.&0 & -\sqrt{n-1} \cr
. & . & .& .&\sqrt{n-1}& 0 
\end{pmatrix} ,
\end{equation} 

and
\begin{equation}
  z =  \frac n\mu
\begin{pmatrix} 
0 & 0 & 0&. & .& . \cr
0 & 0 & 0&. & .&. \cr
0 & 0 & 0&. & .&. \cr
. & . & .&. & .&. \cr
. & . &.& .& 0 &0\cr
. & .& .&. & 0 & 1  
\end{pmatrix}  .
\end{equation}

\noindent
This is the representation which we shall keep in mind
because it is not at the moment clear 
what are representations of (\ref{alg}) for values of $\epsilon$  
between  $\epsilon =0$ and $\epsilon =1$. Matrices $x$ and $y$ are
easily recognizable as quantum-mechanical coordinate and momentum
represented in the Fock basis, of course for $n\to\infty$. The limit
can be consistently imposed at the level of algebra as 
projection to the hyperplane $z=0$. Then the 
commutation relations reduce to
\begin{equation}
 [x,y] = i\kbar .                                             \label{H}
\end{equation} 
Some geometric properties of (\ref{alg})
were analyzed in \cite{trHei}; here we wish to
define gauge fields and therefore 
we need to explore structure of the algebra of 
differential forms. We have established already that
though  algebra (\ref{H}),
defined as a limit of matrix truncations for $n\to\infty$, is 
two-dimensional its cotangent space is three-dimensional.  
In consequence the corresponding geometry differs
from the usual flat geometry of the Moyal space. The same property
is shared by differential calculi on other noncommutative spaces, 
for example  on the fuzzy sphere or on the $\kappa$-Minkowski 
space,~\cite{fuzzy,kappa}. Let us first construct  the exterior 
algebra on the full three-dimensional algebra (\ref{alg}).

 In the approach which we are
using the space of 1-forms $\Omega^1$ is spanned by a preferred set
of forms $ \theta^\alpha $ which is called the frame, \cite{book}.  
Dual to $\theta^\alpha$  are the derivations $e_\alpha$ as 
$\,\theta^\alpha(e_\beta) =\delta^\alpha_\beta$.
Differential $d$ of a function $f$ is defined as 
 $\,df = e_\alpha f \, \theta^\alpha $. The derivations
are inner, generated by momenta $p_\alpha\in \c{A}$,
$\ e_\alpha f = [p_\alpha, f]$. We take $p_\alpha$ to be 
antihermitian, so  for real $f$, $\, e_\alpha f$ is
real too. Since the differential calculus is defined by
$p_\alpha$ it is quite clear that in some respect momenta are
 more fundamental than coordinates. They are also more 
elementary in the sense that formulae look 
simpler when expressed in terms of the momenta. For example, 
it can be shown that $p_\alpha$  always form a quadratic algebra
\begin{equation}
 2P^{\gamma\delta}{}_{\alpha\beta}p_\gamma p_\delta 
- F^\gamma{}_{\alpha\beta} p_\gamma -\frac{1}{\ep} K_{\alpha\beta} =0, \label{2*}
\end{equation} 
that is that commutators between momenta cannot have 
arbitrary form as can commutators between coordinates.
Constants $P^{\gamma\delta}{}_{\alpha\beta}$, $
F^\gamma{}_{\alpha\beta}$ and $ K_{\alpha\beta} $ are 
called the structure elements; defining
\begin{equation}
P^{\gamma\delta}{}_{\alpha\beta} 
=  \frac 12 (\delta^\gamma_\alpha \delta^\delta_\beta -
\delta^\gamma_\beta \delta^\delta_\alpha)
+\ep Q^{\gamma\delta}{}_{\alpha\beta} ,
\end{equation}
we can rewrite (\ref{2*}) as
\begin{equation}
 [p_\alpha,p_\beta] = \frac{1}{\ep}K_{\alpha\beta} 
+F^\gamma{}_{\alpha\beta}p_\gamma 
-2\ep Q^{\gamma\delta}{}_{\alpha\beta}p_\gamma p_\delta .        \label{pp}
\end{equation} 
The choice of the momenta is equivalent to the choice of 
differential calculus and in principle it is not completely fixed.
 For the truncated Heisenberg space we have \cite{trHei}
\begin{equation}
 \epsilon p_1 =i \mu^2 y, \qquad \epsilon p_2 =-i\mu^2 x,\qquad 
\epsilon p_3 =i\mu (\mu z- \frac 12),
\end{equation} 
and the momentum algebra is given by
\begin{eqnarray}
&& [p_1,p_2] =\frac {\mu^2}{2i\epsilon} + \mu p_3,      \nonumber \\[4pt]
&& [p_2,p_3] = \mu p_1 -i\epsilon (p_1p_3+p_3p_1),       \label{palg}\\[4pt]
&& [p_3,p_1] =\mu  p_2  -i\epsilon (p_2p_3 +p_3p_2) .    \nonumber
\end{eqnarray}
One can observe that it has neither the structure of a Lie algebra 
nor the structure of a quantum group.
We identify the structure elements
\begin{equation}
 K_{12} =\frac {\mu^2}{2},\qquad F^1{}_{23} = \mu ,\qquad
 Q^{13}{}_{23} = \frac 12, \qquad  Q^{23}{}_{31} = \frac 12.  \label{struc}
\end{equation}

Since the algebra and the 1-forms commute the algebra
of forms is the tensor product
$\Omega^*({\cal A}) ={\cal A }\otimes \bigwedge^* \ $
of the algebra $\cal A$ and a finite-dimensional exterior
algebra $\,\bigwedge^*\,$ generated by three elements.
Obviously the exterior multplication has to be consistent
if not completely defined by the differential. In the frame formalism
coefficients $ P^{\gamma\delta}{}_{\alpha\beta} \,$
define  the exterior multiplication of two 1-forms $\theta^\alpha$:
\begin{equation}
 \theta^\gamma\theta^\delta 
= P^{\gamma\delta}{}_{\alpha\beta}\theta^\alpha\otimes\theta^\beta, \label{pro}
\end{equation} 
that is
\begin{equation}
\theta^\gamma \theta^\delta +\theta^\delta \theta^\gamma 
= 2 \ep Q^{\gamma\delta}{}_{\alpha\beta} \theta^\alpha \theta^\beta . \label{antico}
\end{equation} 
1-forms $\theta^\alpha$ do not anticommute only when
 the momentum algebra has quadratic terms, $Q^{\gamma\delta}{}_{\alpha\beta} \neq 0 $.
$P^{\alpha\beta}{}_{\gamma\delta}\,$ has to be a projector
\begin{equation}
 P^{\alpha\beta}{}_{\gamma\delta} P^{\gamma\delta}{}_{\eta\zeta} 
= P^{\alpha\beta}{}_{\eta\zeta} ,                            \label{proj}
\end{equation} 
and also
\begin{equation}
 K_{\alpha\beta}P^{\alpha\beta}{}_{\gamma\delta} = K_{\gamma\delta},\qquad
 F^\eta{}_{\alpha\beta}P^{\alpha\beta}{}_{\gamma\delta} = F^\eta{}_{\gamma\delta},\qquad
Q^{\eta\zeta}{}_{\alpha\beta}P^{\alpha\beta}{}_{\gamma\delta} = Q^{\eta\zeta}{}_{\gamma\delta}.                               \label{2}
\end{equation} 
Hermiticity of the basis $\theta^\alpha$ can be assumed as in 
commutative case. Imposing  hermiticity 
on the exterior product implies the condition, \cite{reality},
\begin{equation}
( P^{\alpha\beta}{}_{\gamma\delta})^* P^{\delta\gamma}{}_{\eta\zeta} = P^{\beta\alpha}{}_{\eta\zeta}          .             \label{3}
\end{equation}
As it can be checked, formulae (\ref{proj}) and (\ref{2}) are 
on the truncated Heisenberg space satisfied for all 
values of $\epsilon$,  whereas (\ref{3}) is true for $\epsilon=1$.

The $\ P^{\gamma\delta}{}_{\alpha\beta} \theta^\alpha \theta^\beta\ $ 
form a basis of the space of 2-forms $\Omega^2$. Anticommutation 
relations for truncated  Heisenberg geometry are given by
\begin{eqnarray}
&& (\theta^1)^2 =0,   \quad (\theta^2)^2 =0,  \quad (\theta^3)^2 =0, 
\nonumber\\[4pt]&& 
 \{ \theta^1,\theta^2 \} =0,     
 \label{2forms}\\[4pt]&& 
\{ \theta^1,\theta^3 \} = \ep (  \theta^2\theta^3 -\theta^3\theta^2),
\nonumber \\[-6pt]&&  \nonumber \\[-6pt]&&  
\{ \theta^2,\theta^3 \} = \ep  ( \theta^3\theta^1 -\theta^1\theta^3),       
\nonumber 
\end{eqnarray}
while the canonical basis is
\begin{eqnarray}
&&P^{12}{}_{\gamma\delta}\theta^\gamma\theta^\delta 
=\frac 12\, [\theta^1,\theta^2 ],  \nonumber \\[4pt]
&& P^{13}{}_{\gamma\delta}\theta^\gamma\theta^\delta 
=\frac 12 \,[\theta^1,\theta^3 ] 
+\frac 12 i\epsilon [ \theta^2,\theta^3 ] ,\label{basis} \\[4pt]
&& P^{23}{}_{\gamma\delta}\theta^\gamma\theta^\delta 
=\frac 12 \,[\theta^2,\theta^3 ] 
-\frac 12 i\epsilon [ \theta^1,\theta^3 ] .  \nonumber
\end{eqnarray}
We see that as a basis we can  alternatively use
the set of anticommutators as in commutative 
geometry. In fact, the structure of exterior algebra is almost
completely customary, at least regarding dimensionalities: 
spaces $\Omega^1$ and $\Omega^2$ are three-dimensional, 
while $\Omega^0$ and $\Omega^3$ are  one-dimensional.

Exterior multiplication can be extended to the product of three
1-forms, but an additional constraint has to be fulfilled. 
A necessary condition to define the product uniquely is 
\begin{equation}
\c{C}^{\alpha\beta\gamma}{}_{\eta\zeta \xi}\,
\theta^\eta \theta^\zeta \theta^\xi = 0  ,         \label{YB}
\end{equation} 
\vskip1pt \noindent
where $\c{C}^{\alpha\beta\gamma}{}_{ \eta\zeta\xi} 
= \c{C}^{\beta\gamma}{}_{\delta\epsilon} 
\c{C}^{\alpha \delta}{}_{\eta\lambda} C^{\lambda\epsilon}{}_{\zeta\xi} 
-\c{C}^{\alpha\beta}{}_{\delta\epsilon} 
\c{C}^{\epsilon\gamma}{}_{\lambda\xi} \c{C}^{\delta\lambda}{}_{\eta\zeta} \,
$ and constants
\begin{equation}
\c{C}^{\gamma\delta}{}_{\alpha\beta}
= \delta^\gamma_\alpha \delta^\delta_\beta - 2  P^{\gamma\delta}{}_{\alpha\beta} 
\end{equation}
define the operation which reverses the order of indices in the exterior product,
\begin{equation}
\theta^\alpha \theta^\beta 
= -\c{C}^{\alpha\beta}{}_{\gamma\delta} \theta^\gamma \theta^\delta. 
\end{equation} 
Equation (\ref{YB}) is a weak form of the braid relation and it is 
satisfied in our case. This means that the product of three
1-forms can be defined unambiguously\footnote{Relations 
like (\ref{proj}), (\ref{3}), (\ref{YB}) which include
 structure elements  were checked using {\sl Mathematica}.}.

 What one needs in calculation is the twisted-antisymmetric tensor,
a generalization of the usual 
 $\, \delta^{\alpha\beta\gamma}_{\zeta\eta\xi} $
(which in three dimensions reduces to the product, 
$\delta^{\alpha\beta\gamma}_{\zeta\eta\xi} 
=\epsilon^{\alpha\beta\gamma} \epsilon_{\zeta\eta\xi}$).
The neeed generalization is given by
\begin{equation}
 \Delta^{\alpha\beta\gamma}_{\zeta\eta\xi}
=\tfrac 13\, (-\delta^\alpha_\zeta P^{\beta\gamma}{}_{\eta\xi}
- \c{C}^{\alpha\beta}{}_{\zeta\rho}P^{\rho\gamma}{}_{\eta\xi} +
\c{C}^{\beta\gamma}{}_{\rho\sigma} \c{C}^{\alpha\rho}{}_{\zeta\tau} 
P^{\tau\sigma}{}_{\eta\xi} ) . 
\end{equation} 
Unlike $\,\delta^{\alpha\beta\gamma}_{\zeta\eta\xi}\, $,  
noncommutative $\,\Delta^{\alpha\beta\gamma}_{\zeta\eta\xi} \,$ 
is not a projector; however on
the product of three 1-forms it acts as one,
\begin{equation}
\Delta^{\alpha\beta\gamma}_{\zeta\eta\xi} \theta^\zeta \theta^\eta \theta^\xi 
= \theta^\alpha \theta^\beta \theta^\gamma .
\end{equation} 
We shall need $\,\Delta^{\alpha\beta\gamma}_{\zeta\eta\xi}\, $ 
to define the volume 3-form $\Theta\,$, and later on, the action. 
By definition the integral
of a 3-form  $\alpha = f \Theta\ $ is given by
 $\, \int\alpha = \tr f $.

It is not difficult to write the algebra 
of 3-forms on truncated Heisenberg space  explicitly. 
From (\ref{2forms}) and the associativity 
of the exterior product we obtain
\begin{eqnarray}
&&
\theta^1\theta^3\theta^1 =\theta^2\theta^3\theta^2, \nonumber \\[4pt]
&& \theta^1\theta^2\theta^3 = -\theta^2\theta^1\theta^3 =
\theta^3\theta^1\theta^2 = -\theta^3\theta^2\theta^1
= i\,\frac{\epsilon^2 -1}{2\epsilon}\,\theta^2\theta^3\theta^2 ,
\label{3forms} \\
&&\theta^1\theta^3\theta^2 = -\theta^2\theta^3\theta^1 
= i\,\frac{\epsilon^2 +1}{2\epsilon}\, \theta^2\theta^3\theta^2  . \nonumber
\\[4pt]
&& \theta^3\theta^1\theta^3 =0, \qquad 
\theta^3\theta^2\theta^3 =0.\nonumber
\end{eqnarray}
There is obviously only one independent 3-form, that is 
the volume form is unique. We  define it as
\begin{equation}
\Theta = - \frac{i}{2\epsilon} \,\theta^2\theta^3\theta^2 
\end{equation}
in order that it reduce to $\,\theta^1\theta^2\theta^3 \,$ in the
commutative limit.  Note that the product of three 1-forms is not
cyclic, for example $ \theta^1\theta^3\theta^2 \neq \theta^3\theta^2\theta^1 $.
  Relations (\ref{3forms}) can be rewritten as
\begin{equation}
\begin{array}{l}
[\theta^1,\theta^2]\,\theta^3 = \theta^3
 [\theta^1,\theta^2] = 2(1 -\epsilon^2 )\Theta,
\\[4pt]
[\theta^2,\theta^3] \,\theta^1 = \theta^1
[\theta^2,\theta^3] = 2\Theta,
\\[4pt]
[\theta^2,\theta^3]\,\theta^2 = -\theta^2[\theta^2,\theta^3] 
=2 i \epsilon \Theta,
 \\[4pt]
[\theta^3,\theta^1]\,\theta^2 = \theta^2 [\theta^3,\theta^1] = 2\Theta ,
 \\[4pt] 
[\theta^3,\theta^1]\,\theta^1 = -\theta^1[\theta^3,\theta^1] = -2i \epsilon \Theta .
\end{array}                                             \label{3forms1}
\end{equation} 
From~(\ref{3forms}) we see that the value $\,\epsilon =1$ 
is special: it gives for example $\,\theta^1\theta^2\theta^3 =0 $, which 
is in the commutative case unusual. Nonetheless the algebra of 3-forms 
is nondegenerate with nonvanishing elements 
$\,\theta^1\theta^3\theta^1$, $\,\theta^2\theta^3\theta^2$,
$\,\theta^1\theta^3\theta^2$ and $\,\theta^2\theta^3\theta^1 $. 
As we have mentioned value $\epsilon =1$ is important from the 
point of view of  representations. Therefore we wish to include it 
explicitly, so we define the Hodge-dual as
\begin{equation}
^*\left(\tfrac 12\,[\theta^1,\theta^2]\right) =  \theta^3, \qquad 
^*\left(\tfrac 12\,[\theta^2,\theta^3]\right) = \theta^1 , \qquad 
^*\left(\tfrac 12\,[\theta^3,\theta^1]\right) = \theta^2 .  \label{Hodge}
\end{equation} 
This definition modifies the usual normalization, giving for example
\begin{equation}
^*\left(\tfrac 12\,[\theta^1,\theta^2]\right)\, 
\tfrac 12\, [\theta^1 ,\theta^2] =(1 -\epsilon^2 )\, \Theta . \label{Hodgenorm}
\end{equation}
This implies in particular that for $\epsilon =1$ the 2-form  
$\,[\theta^1,\theta^2] \,$ does not have a Hodge dual, 
and consequently the corresponding term will be absent 
from the Yang-Mills lagrangian. Note that relations (\ref{Hodge}) 
give the Hodge dual uniquely only if, when multiplying forms with their 
duals the  products are symmetrized.

Geometric characteristics of the truncated Heisenberg space as the
connection and the curvature were discussed in \cite{trHei}.
In order to define gauge fields  we only need the
Ricci rotation coefficients $C^\alpha{}_{\beta\gamma}\, $, 
$\,  d\theta^\alpha = -\tfrac 12 C^\alpha{}_{\beta\gamma}
\theta^\beta\theta^\gamma \,$.
They are determined by the structure elements,
\begin{equation}
C^\gamma{}_{\alpha\beta} = F^\gamma{}_{\alpha\beta} -4\ep
Q^{\gamma\delta}{}_{\alpha\beta}p_\delta .
\end{equation}
For the truncated Heisenberg space we have
\begin{eqnarray}
&& C^1{}_{23} =- C^1{}_{32} = 2\mu^2 z ,\quad 
C^2{}_{31} =- C^2{}_{31} =2\mu^2 z ,\quad 
C^3{}_{12} =- C^3{}_{21} =\mu ,\nonumber \\[4pt]
&& C^3{}_{13} =- C^3{}_{31} = 2\mu^2 x , \quad
C^3{}_{23} =- C^3{}_{32} = 2\mu^2 y  .
\end{eqnarray}

\initiate
\section{Gauge fields}

As mentioned already, to define gauge symmetries
we use representation-independent formulation of \cite{gauge}
which we will slightly generalize. The focus will be
mostly on  formulae and their application to
the truncated Heisenberg geometry; for more mathematical recent
reviews see \cite{de Goursac:2007gq} or \cite{Cagnache:2008tz}.
The gauge potential $A$ and the field strength $F$ are respectively
a 1-form and a 2-form,
 \begin{equation}
 A = A_\alpha\theta^\alpha,\qquad F = dA +A^2 = \tfrac 12 \, 
F_{\alpha\beta} \theta^\alpha \theta^\beta .                                                  \label{AF}
\end{equation} 
They are antihermitian and dimensionless; 
of course,  $A$ and $F$ are functions of noncommutative coordinates.
As the gauge group we take  noncommutative $U_1$, the group of 
all unitary elements of $\c{A}$; the group elements are denoted by $g$.  
In the finite-matrix representation $U_1$ consists of 
all unitary $n\times n$ matrices and thus, as the set of elements 
with group multiplication, the noncommutative 
$U_1$ is equal to the usual  $U_n$ 
which acts on fields on a commutative space.

$A$ is a connection so it 
transforms as $ A^\prime =  g^{-1} A g + g^{-1} dg $, that is
\begin{equation}
  A^\prime_\alpha  =g^{-1}  A_\alpha g + g^{-1} e_\alpha g .
\end{equation}
The field strength transforms in the adjoint representation, 
$F^\prime = g^{-1} Fg $. $F$ is a 2-form so it can be expanded 
in the basis (\ref{pro}). From (\ref{proj}) we see that components
$F_{\alpha\beta}$ satisfy
\begin{equation}
 F_{\zeta\eta} = F_{\alpha\beta}P^{\alpha\beta}{}_{\zeta\eta} .
\end{equation} 
This means that the components of the field strength are 
in our case antisymmetric, because coefficients 
$P^{\alpha\beta}{}_{\zeta\eta} \, $ are antisymmetric 
in the lower pair of indices. 
From definition  (\ref{AF}) we obtain
\begin{equation}
  dA + A^2  = (e_\beta A_\gamma -\tfrac 12 A_\alpha C^\alpha{}_{\beta\gamma}+ A_\beta A_\gamma )P^{\beta\gamma}{}_{\zeta\eta}\theta^\zeta \theta^\eta ,
\end{equation}
and the field strength is given by
\begin{equation}
 F_{\zeta\eta} =e_{[\zeta} A_{\eta]} - A_\alpha C^\alpha{}_{\zeta\eta}+[A_\zeta, A_\eta] + 2i\epsilon (e_\beta A_\gamma)Q^{\beta\gamma}{}_{\zeta\eta}+ 2i\epsilon A_\beta A_\gamma Q^{\beta\gamma}{}_{\zeta\eta} .                   \label{F}
\end{equation}
In case of vanishing torsion, that is when 
$\ \omega^\alpha{}_{[\beta\gamma]} = C^\alpha{}_{\beta\gamma}  $,
(\ref{F}) can be written as
\begin{equation}
  F_{\zeta\eta} =\nabla_{[\zeta} A_{\eta]} +[A_\zeta, A_\eta] + 2i\epsilon (e_\beta A_\gamma)Q^{\beta\gamma}{}_{\zeta\eta}+ 2i \epsilon A_\beta A_\gamma Q^{\beta\gamma}{}_{\zeta\eta} ,
\end{equation}
where the expression
\begin{equation}
\nabla_\zeta A_\eta = e_\zeta A_\eta -A_\alpha \omega^\alpha{}_{\zeta\eta}                                                                         \label{cov}
\end{equation} 
denotes the gravity-covariant derivative of the vector $A_\alpha$.

An important property of noncommutative spaces with inner 
derivation-based calculus is the existence of a preferred connection $\theta$,
\begin{equation}
 \theta = -p_\alpha \theta^\alpha  .
\end{equation} 
The differential can be expressed in the form $  df = -[\theta,f]. $
As one can show easily,
\begin{equation}
 d\theta + \theta^2 = \frac{1}{2\ep} K_{\alpha\beta}\theta^\alpha \theta^\beta  .
                                                                  \label{dth+th^2}
\end{equation} 
 $\theta \,$ is an example of the Dirac operator
in the sense of Connes, 
\cite{Diracop}. It is invariant under the action of the gauge group: 
one can see it from
\begin{equation}
 \theta^\prime = g^{-1}\theta g + g^{-1}dg ,
\end{equation} 
and
\begin{equation}
 g^{-1}\theta g = g^{-1}[\theta ,g] +g^{-1} g\theta = -g^{-1}dg +\theta .
\end{equation} 
The difference between  connections $A$ and $\theta$,
$\, \tX = A-\theta $,
transforms in the adjoint representation; 
coefficients $ \tX_\alpha = p_\alpha + A_\alpha\ $ are  called the
 covariant coordinates. 
Expressing the field strength in terms of $\tX_\alpha$ and the 
structure elements we obtain
\begin{equation}
 F_{\alpha\beta} = 2P^{\gamma\delta}{}_{\alpha\beta}{\tX}_\gamma {\tX}_\delta -F^\gamma{}_{\alpha\beta}{\tX}_\gamma -\frac{1}{i\epsilon}K_{\alpha\beta} ,                         \label{Falbe}
\end{equation} 
that is 
\begin{equation}
F = \tX^2  - \frac 12 F^\gamma{}_{\alpha\beta}{\tX}_\gamma \theta^\alpha\theta^\beta -\frac{1}{2i\epsilon}K_{\alpha\beta} \theta^\alpha\theta^\beta     .\label{FX}
\end{equation} 
Covariant coordinates diverge in the commutative limit because $p_\alpha$ 
do (in our case for example, $ p_1 =  \frac{i\mu^2}{\epsilon} y$, 
$ p_2 =  -\frac{i\mu^2}{\epsilon} x$, etc. so for $\epsilon\to 0$, 
$p_\alpha\to\infty$). For quantization  it is thus better to 
express the action in $A_\alpha$, $F_{\alpha\beta}\,$ as
then we have the control of the commutative limit.
On the other hand, if we write the lagrangian in terms of 
$\tX_\alpha$ we obtain a polynomial, that is, the kinetic 
term is absent. The corresponding theory is  equivalent to 
a matrix model, \cite{matrixmod}, for which the classical equations of 
motion  can often be solved. Covariant coordinates are very useful to 
keep track of transformation properties under the symmetry group 
and therefore we will usually write equations parelelly in  
$A$ and in $\tX$.

To distinguish the values of the gauge field on 
the full three-dimensional truncated Heisenberg space from 
those  defined  intrinsically on the 
two-dimensional Moyal plane, we denote the 
former by $\tA_\alpha, \ \tF_{\alpha\beta}$ ($\alpha,\beta = 1,2,3$)
and the latter by
$A_\alpha, \ F_{\alpha\beta}$ ($\alpha,\beta = 1,2$). 
We will be interested in the subspace $z=0\,$, where 
$\,p_3 =-\frac{i\mu}{2\epsilon}\, $, $\,e_3=0\,$. On this 
subspace the  component $\tA_3$ transforms as a scalar field 
in the adjoint representation, 
$\,\tA^\prime_3 = g^{-1} \tA_3 g $. We denote 
\begin{equation}
\tA_3 = \phi,\quad \tA_1 = A_1,\quad \tA_2=A_2 ,
\end{equation} 
and equivalently,
\begin{equation}
 \tX_1 = p_1+A_1,\quad \tX_2=p_2+A_2 ,   \quad\tX_3 = -\frac{i\mu}{2\epsilon} +\phi .
\end{equation}
On the Moyal plane we would have 
\begin{eqnarray}
&&{D}_\alpha \phi = [p_\alpha + A_\alpha ,\phi] = e_\alpha \phi+  [A_\alpha, \phi] ,\\[4pt]
&&F_{12} = e_1A_2 -e_2 A_1 + [A_1, A_2] , \nonumber
\end{eqnarray} 
while from (\ref{F}) on the truncated Heisenberg space 
\begin{eqnarray}
&&\tF_{12} = e_1\tA_2 -e_2 \tA_1 +[\tA_1, \tA_2] -\mu \tA_3 ,\nonumber \\[6pt]
&&\tF_{13} =[p_1+\tA_1, \tA_3] - i\epsilon \{ p_2 +\tA_2, \tA_3\} +2\mu^2\tA_2 z , 
                                                       \label{3.13}\\[6pt]
&& \tF_{23} =[p_2+\tA_2, \tA_3] + i\epsilon \{ p_1 +\tA_1, \tA_3\}-2\mu^2 \tA_1 z\nonumber .
\end{eqnarray}
In particular, for $z=0$ we obtain
\begin{eqnarray}
&&\tF_{12} = F_{12}-\mu \phi =[\tX_1,\tX_2] + \frac{i\mu^2}{\epsilon} -\mu\phi ,               \nonumber  \\[6pt]
&&\tF_{13} ={D}_1\phi - i\epsilon \{ p_2 + A_2, \phi\} = [\tX_1,\phi] -i\epsilon \{ \tX_2, \phi\} , \label{field} \\[6pt]
&&\tF_{23} ={D}_2\phi + i\epsilon \{ p_1 +A_1, \phi\} = [\tX_2,\phi] +i\epsilon 
\{ \tX_1, \phi \}\nonumber .
\end{eqnarray}
Clearly, on the given subspace vector potential $\tA_\alpha$ splits
into a scalar mode $\phi$ and a vector mode $A_\alpha$. 
We mentioned earlier that the gauge field on the fuzzy sphere 
behaves similarly, only in that case the scalar component
corresponds to the radial degree of freedom $\phi =x^iA_i +A_i x^i$
while the remaining two gauge degrees of freedom are tangential, 
\cite{Grosse:1992bm}; see also \cite{matrixmod,Kimura}.
The  difference in dimensionsionalities of the basic and the cotangent 
space, apparently somewhat counterintuitive, seems to 
follow naturally from the relation which  noncommutative geometry has with 
d-brane physics, \cite{Alekseev:2000fd}. The same kind of effect
in the usual Kaluza-Klein reduction one obtains from the assumption
that the fields do not depend on the internal that is additional coordinates
($z$ in our case, $r$ on the sphere).

\initiate
\section{Yang-Mills action}

We showed that the exterior algebra on the truncated Heisenberg 
space admits a unique volume form and therefore the 
integration is well defined. To obtain the Yang-Mills action 
we start from a three-dimensional expression.
The third `integral' will be absorbed implicitly in a
rescaling of the coupling constant after reduction to $z=0$.
The Yang-Mills action is given by
\begin{equation}
 \cS_{YM} =   \frac{1}{16} \,\tr   (\tF ^*\tF+ ^*\tF \tF).
                                                               \label{ym}
\end{equation} 
As it was mentioned,
we have to symmetrize the product of forms.
For $ \epsilon =0$,  (\ref{ym})  reduces to the standard expression
$ \ \cS_{YM} =  \frac{1}{4} \,\tr \, \tF_{\alpha\beta}\tF^{\alpha\beta} \, $.
In our case due to normalization (\ref{Hodge}) we have
\begin{equation}
 \cS_{YM} =  \frac{1}{2}\, \tr \, \Big( (1-\epsilon^2)\,\tF_{12}\tF^{12} +
\tF_{13}\tF^{13} +\tF_{23}\tF^{23}\Big)   .
\end{equation} 
Introducing expressions (\ref{field}) for the first term we obtain
\begin{equation}
 \tr(\tF_{12})^2 = \tr\Big( (F_{12})^2 -2\mu F_{12} \, \phi  + \mu^2\phi^2\Big) ,
\nonumber
\end{equation} 
while the sum of the other two terms after various simplifications becomes
\begin{eqnarray}
 &&\tr \Big( (\tF_{13})^2 +(\tF_{23})^2\Big)  =\tr \Big( (D_1 \phi)^2 +(D_2 \phi)^2 + 4 \mu ^2 \phi^2 + 4i\epsilon F_{12}  \phi^2     \nonumber \\[4pt]
&&\phantom{(F_{13})^2 +(F_{23})^2 = \tr\qquad \ } -\epsilon^2\{ p_1 +A_1, \phi\}^2 
-\epsilon^2\{ p_2 +A_2, \phi\}^2 \Big) .\nonumber
\end{eqnarray}
Therefore the Yang-Mills action is 
\begin{eqnarray}
 &&\cS_{YM}= \frac 12 \,\tr \Big( (1-\epsilon^2)(F_{12})^2 -2 (1-\epsilon^2)\mu F_{12}\phi + (5 -\epsilon^2) \mu^2 \phi^2 + 4i\epsilon F_{12}\phi^2  \label{L}
\\[4pt]
&&\phantom{S = \frac 12 \tr \quad\ } + (D_1 \phi)^2 +(D_2 \phi)^2 -\epsilon^2\{ p_1 +A_1, \phi\}^2 -\epsilon^2\{ p_2 +A_2, \phi\}^2 \Big)  , \nonumber
\end{eqnarray}
or using the covariant coordinates, 
\begin{eqnarray}
 &&\cS_{YM}= \frac 12 \, \tr \Big( (1-\epsilon^2) \big([\tX_1,\tX_2]^2+\mu^2\phi^2-\frac{2i\mu^3}{\epsilon}\phi -2\mu\,[\tX_1,\tX_2]\,\phi
\big) \\[4pt]
&&\phantom{\cS_{YM}= \tfrac 12 \tr }
+ 4\ep \,[\tX_1,\tX_2]\,\phi^2 +[\tX_1,\phi]^2 +[\tX_2,\phi]^2 -\epsilon^2\{ \tX_1,\phi\}^2
-\epsilon^2\{ \tX_2,\phi\}^2  \Big)\nonumber .
\end{eqnarray} 
The action is obviously gauge invariant.

One observes immediately that for $\epsilon =1$ the kinetic term
for the gauge field $F_{12}$ is absent: the action is almost identical to the 
induced gauge action obtained in \cite{de Goursac:2007gq,Grosse:2007dm} by 
the path integration of the scalar field. The difference is in the additional 
terms which mix $F_{12}$ and $\phi$: the kinetic term $  {F}_{12}\phi$
and the interaction term $ {F}_{12}\phi^2$. 
Indeed this difference is significant as it changes the vacuum.
We can see it from the equations of motion:
\begin{eqnarray}
&& \frac{\delta \cS_{YM} }{\delta\phi} = 0 =- (1-\epsilon^2)\,\mu F_{12} 
+(5-\epsilon^2)\mu ^2\phi + 2i\epsilon \{ F_{12},\phi \} \label{45}\\[4pt]
&& \phantom{\frac{\delta \cS_{YM} }{\delta\phi} = 0 = }
- D^\alpha D_\alpha \phi -\epsilon^2\{ p^\alpha + A^\alpha,
\{ p_\alpha + A_\alpha ,\phi \} \}  ,
\nonumber\\
 && \frac{\delta \cS_{YM} }{\delta A_\alpha} = 0 =(1-\epsilon^2)\, \epsilon^{\alpha\beta}D_\beta ( F_{12}-\mu \phi) + 2i\epsilon \epsilon^{\alpha\beta} 
\{ D_\beta \phi ,\phi \} \label{46} \\[4pt]
&&\phantom{ \frac{\delta \cS_{YM} }{\delta A_\alpha} =0 =}
-[D_\alpha \phi,\phi ] -\epsilon^2\{ \{ p^\alpha +A^\alpha, \phi \}, \phi \} .\nonumber
\end{eqnarray} 
It is difficult to solve these equations in the most general case.
Confining to  constant solutions we obtain two,
\begin{equation}
 A_1 =0,\ A_2 =0,\ \phi =0 \qquad {\rm and}\qquad 
\tX_1 =0,\ \tX_2 =0,\ \phi =\frac{i\mu}{\epsilon} .          \label{vac}
\end{equation}  
Obviously the first solution is the usual vacuum which can be
used for quantization. The second solution $\,A_1 =-\tfrac{i\mu^2}{\epsilon}\, y$, 
$\,A_2 =\tfrac{i\mu^2}{\epsilon}\,x$, $\,\phi =\tfrac{i\mu}{\epsilon} \,$, has a
 constant  field strength
$\, F_{12} = \tfrac{i\mu^2}{\epsilon}$, and presumably nonminimal energy.

\initiate
\section{Chern-Simons action}

It is not completely straightforward to deduce
what would in general be the noncommutative equivalent of 
the usual Chern-Simons action, \cite{chern}. Clearly it has to be 
an expression which is, up to surface terms, invariant under 
the gauge group 
\begin{equation}
 \delta \cS_{CS} =\delta \tr L_{CS} = 0. \label{sl}
\end{equation}
One would impose further the correct 
commutative limit. However, not in all cases a 
topological definition would be appropriate
because considerations which include surface terms
are in the case of finite matrix spaces vacuous: these 
spaces do not have boundary (that is, $\tr df $ is always 
zero because $df$ is expressed in terms of commutators). 
Also, it is not clear that it is in general possible to solve 
Equation~(\ref{sl}) and find the current
\begin{equation}
 \delta L_{CS} = dJ
\end{equation} 
as in the commutative case. 
Some kind of  `canonical' differential 
calculus,  analogous to the de Rham calculus and
defined in any number of dimensions does not exist for noncommutative 
spaces; therefore in formulae it is not justified
{\it a priori} to pass from space of 
one dimension to the space of another. 

In  \cite{p},
as generalization of the Chern-Simons action
for the space with constant noncommutativity
Polychronakos proposed the following action
\begin{equation}
\cS_{CS} = \alpha  \,\tr \, \tX^{2n+1} \,    ,                  \label{cs}
\end{equation}
where $\tX$ is the covariant coordinate 1-form,
 $\tX = \tX_\alpha \theta^\alpha $.
Clearly, as $\tX$ transforms in the adjoint
representation, (\ref{cs}) is invariant under the gauge 
group and contains terms  of the correct order, for 
example  the usual  $F^n A$. Moreover, 
(\ref{cs}) has a reasonable commutative limit: the
limit of the  noncommutative $U_1$  action (\ref{cs})  
is the nonabelian 
$U_n$ Chern-Simons action defined on  commutative space, \cite{p}.

The advantage of action $\tr \tX^3$
in our framework  is that 
it explicitly symmetrizes the factors of the volume 3-form $\Theta$ which  are
 otherwise not cyclic. We will therefore  use (\ref{cs}) as a
definition of the Chern-Simons action and explore its implications
 for the truncated Heisenberg space. Applying the projector
$\,\Delta^{\alpha\beta\gamma}_{\zeta\eta\xi}\, $ we have
\begin{equation}
 \tr\tX^3 = \tr \left( \tX_{\alpha}\tX_{\beta}\tX_\gamma\,  
\Delta^{\alpha\beta\gamma}_{\zeta\eta\xi}\, \theta^\zeta \theta^\eta \theta^\xi\right),
\end{equation}  
and therefore  we obtain, in components,
\begin{equation} 
\cS_{CS} = \frac{\alpha\mu}{3}\, \tr \Big((3-\epsilon^2)\,[\tX_1,\tX_2]\,\tX_3 
+ 2i\epsilon \,(\tX_1^2 +\tX_2^2)\,\tX_3 \Big) .
\end{equation} 
Equivalently in terms of the gauge potentials,
\begin{equation}
 \cS_{CS} =  \frac{\alpha\mu}{3}\, \tr\, 
\Big( (3-\epsilon^2)(F_{12}-\frac{i\mu^2}{\epsilon})\phi 
+\frac{2\ep}{3}\big( (p_1+A_1)^2 + (p_2 + A_2)^2\big) (\phi -\frac{i\mu}{2\epsilon})
\Big) .
\end{equation} 
The Chern-Simons action  also depends on coordinates.
Its variations  are
\begin{eqnarray}
&& \frac{\delta \cS_{CS} }{\delta\phi} = \frac{\alpha\mu}{3} \big( (3-\epsilon^2)(F_{12}-\frac{i\mu^2}{\epsilon})+2\ep\big( (p_1+A_1)^2 + (p_2 + A_2)^2 \big) \big) ,  \\[8pt]
&& \frac{\delta \cS_{CS} }{\delta A_\alpha} =  \frac{\alpha\mu}{3} \big( (3-\epsilon^2)\epsilon^{\alpha\beta} D_\beta\phi +2\ep \{ p_\alpha + A_\alpha, \phi -
\frac{i\mu}{2\epsilon} \}  \big)           .
\end{eqnarray} 
The equations which correspond to the pure Chern-Simons action have one 
constant solution, $\tX_1 =0$, $\tX_2 =0$ , 
$\phi = \frac{i\mu}{\epsilon}$. The other vacuum $A_1=0$, $A_2=0$, $\phi =0$ 
of (\ref{45}-\ref{46}) is  absent in general. The 
sum $\,\cS_{YM}+\cS_{CS}\, $  however has interesting 
properties: by an appropriate choice of  coefficient $\alpha$,
the mixed term $F_{12}\phi$ (which is potentially difficult for quantization) 
can be cancelled. Furthermore, the sum of the two  actions
has  constant solution $A_1 =0$, $A_2=0$, $\phi = i\mu$ (describing
the spontaneous symmetry breaking) for 
a particular choice of parameters $\epsilon =1$, $\alpha =6$. 
Whether one should include $ \cS_{CS}$ in the gauge field 
action or not is not completely clear; defining the BRST complex
we shall proceed with $ \cS_{YM}$ only, the inclusion 
of $ \cS_{CS}$ being straightforward.

\initiate
\section{BRST invariance}

The gauge fixing of the Yang-Mills action can be done 
straightforwardly. The simplest choice is the Lorentz gauge, 
\begin{equation}
 \cG = e_\alpha A^\alpha =\p_1 A^1 +\p_2 A^2.                    \label{G}
\end{equation} 
According to the usual procedure the quantum action is given by
\begin{equation}
 \cS=\cS_{YM} + \cS_{gf}
\end{equation} 
with
\begin{equation}
 \cS_{gf} = 
\tr \left( B e_\alpha A^\alpha +\frac{\alpha}{2} B B -\bar c e_\alpha D^\alpha c \right) ,
\end{equation} 
where we introduced the ghost $c$, the antighost $\bar c$ and the
auxilliary field $B$. The BRST transformation $s$ acting on the gauge 
potential can be defined as
\begin{equation}
  sA_\alpha = D_\alpha c = e_\alpha c + ig[A_\alpha, c].
\end{equation} 
 ${F}_{\alpha\beta}$, $\phi$ and $\tX_\alpha = p_\alpha+A_\alpha $  
transform in the adjoint representation and therefore we have
\begin{eqnarray}
&&
s{F} _{\alpha\beta} = [{F}_{\alpha\beta}, c] , \\[4pt]
&& s\phi =[\phi,c] , \\[4pt]
&& s(\tX_\alpha) =[\tX_\alpha, c] =e_\alpha c+ig[A_\alpha,c] = sA_\alpha .
\end{eqnarray}
This means also that the momenta $p_\alpha$ are BRST-invariant, $sp_\alpha = 0 $.
Of course the Leibniz rule for $s$ holds, for example
\begin{equation}
s\{\tX_\alpha,\phi\} =\{ [\tX_\alpha,c\,],\phi\} +\{\tX_\alpha,[\phi,c\,]\}
 = [\{\tX_\alpha,\phi\},c\,] .
\end{equation}
For the ghosts and the auxiliary field we can also define the BRST
transformation as usual 
\begin{equation}
 s c = -c^2 ,\qquad
s\bar c =B , \qquad
sB =0 .
\end{equation} 
With these definitions it is easy to check that $s$ is nilpotent,
$s^2 =0$, and then also that $s\cS_{gf}=0$ as the gauge-fixing term 
is equal to
\begin{equation}
 \cS_{gf} = \tr s(\bar c e_\alpha A^\alpha +\frac{\alpha}{2} \bar c B) .
\end{equation} 
For the  classical Yang-Mills part of the action 
$s\cS_{YM}=0$ because of its gauge invariance.

It is quite clear that the gauge fixing (\ref{G}) is a possible gauge 
choice. However as we are in the curved space, one might wonder whether 
a more natural choice is the covariant gauge,
$ \nabla_\alpha A^\alpha$. The covariant derivative was defined 
already in (\ref{cov}); we denote
\begin{equation}
 \cF =e_\alpha A^\alpha +A^\beta \omega^\alpha{}_{\alpha\beta} .
\end{equation} 
To calculate $\cF$ we need the linear connection. 
Asssuming that 
$\omega_{\alpha\beta\gamma} =\frac 12 ( C_{\alpha\beta\gamma} - C_{\beta\gamma\alpha} +C_{\gamma\alpha\beta})$ as in  \cite{trHei},
we obtain 
\begin{equation}
 \omega^{\alpha}{}_{\alpha 1} = -2\mu^2 x,\quad
\omega^{\alpha}{}_{\alpha 2} = -2\mu^2 y,\quad 
\omega^{\alpha}{}_{\alpha 3} =0,
\end{equation} 
 and therefore
\begin{equation}
 \cF = \p_1 A_1 + \p_2 A_2 -2 \mu^2  A_1 x -2 \mu^2A_2  y .
\end{equation}
As $\cF$  is not hermitian  the possible
gauge choices are  $\cF\cF^\dagger$  or  
$\frac 14 (\cF + \cF^\dagger)^2$.
In both cases the procedure to introduce $\cS_{gf}$ and 
prove the BRST invariance is straightforward, only the 
corresponding expressions are somewhat longer.
To the comparative advantages and shortcomings of 
different gauge choices we shall return in our future work.

\initiate
\section{Conclusions}

The aim of the paper was to derive the action
 for the $U_1$ gauge field on the truncated Heisenberg 
algebra and to reduce it to  the Moyal subspace, in order 
to obtain a candidate for renormalizable noncommutative 
gauge theory. Various other proposals were discussed
in the literature, \cite{de Goursac:2007gq,Grosse:2007dm,
Blaschke:2007vc,Blaschke:2009hp,ren}. Our main
idea was to apply the geometric logic developed in
\cite{trHei}: it was shown there that 
renormalizability of the  Grosse-Wulkenhaar action 
can be attributed to the interaction with 
the background curvature. If indeed geometric properties 
of the background space play a role, a similar result
is to be expected for the gauge and other fields. 
An alternative way to understand how renormalizability can
be related to a noncommutative geometry is the fact that 
the underlying space in question is an algebra of finite matrices.
Defining the scalar or the gauge fields on this algebra 
provides therefore a geometrically consistent way 
to define matrix regularization.

The truncated Heisenberg algebra is  a 
three-di\-men\-si\-onal
noncommutative space. Properties of its cotangent space 
 were discussed in  \cite{trHei}; here
in order to define gauge fields we explore
the structure of the spaces of 2-forms and
 3-forms. This enables to define the 
Hodge-dual and the volume form
and consequently to obtain the Yang-Mills action.
When written in frame components the gauge potentials 
couple to the  connection, which results in an
explicit coordinate dependence of the 
lagrangian. This is the property we aimed to obtain, as
in the case of the scalar field the coordinate dependence
modified the usual propagator to the Mehler kernel
and that had ensured the absence of the UV/IR mixing 
and eventually, the renormalizability.

To get a theory in two dimensions\ in the second
step of the construction we constrain 
to the subspace $z=0\,$; this subspace
 is algebraically but not geometrically equivalent to the 
Moyal space.  The degrees of freedom of the
vector potential $\,\tA_\alpha$, $\alpha = 1,2,3\,$  split
then into a scalar field  $\phi$ and a two-dimensional 
gauge field $A_\alpha$,  $\alpha = 1,2$. $\,\phi$ and
$A_\alpha$ are coupled; the coupling is completely
fixed because it comes from a
higher-dimensional Yang-Mills action. It
is possible to write the resulting lagrangian 
in terms of the covariant coordinates also. The corresponding
classical equations of motion have a solution
$\, \phi =0$, $A_\alpha =0$ suitable 
as a vacuum for the quantization. 
We  also discuss the Chern-Simons action
and show how it changes the vacuum. Finally, we 
introduce the appropriate gauge fixing and prove
the BRST invariance of the gauge-fixed action.

The coordinate dependence of the lagrangian was in some previous works 
attributed to the external electromagnetic field, \cite{ls}; 
here we relate it to the background gravity. 
Note that there is a further possibility to include coordinates 
covariantly, which is due to a special form of the differential 
calculus, $df  = -[p_\alpha ,f]\theta^\alpha =-[\theta,f]$.
The coordinate-dependent quantities
$\,\tX_\alpha = \tA_\alpha +p_\alpha\,$ in this case
 transform in the adjoint representation of the gauge group
and obviously they can be included in
the action invariantly, \cite{Cagnache:2008tz}. 

Let us elaborate on relations between the gravitational and the
gauge fields on noncommutative spaces a bit further.
It has recently been noticed~\cite{Ste} that one can interpret 
noncommutative
gravity as an induced theory obtained by summing over a set of matrix
models of electromagnetism in a way which is reminiscent of the
`induction' of (euclidean) gravity~\cite{Sakharov:1967pk} by 
summing over all
the quantum fluctuations of a scalar field. One can give a
`derivation' of this result in three steps. One first notices that the
Dirac operator $\theta$ can be interpreted as an electromagnetic
potential. One then recalls that this same operator determines the
differential calculus. Finally one reiterates the argument that the 
differential calculi over an algebra stand in one to one correspondence 
with the metrics consistent with the algebra. Some details of 
the first step are given in Section~3; a  discussion of the other
steps can be found in the literature, \cite{book}.

To summarize: we obtained a geometric action which contains
besides the gauge field a scalar. Their coupling is  of
a particular form and in fact the whole construction has similarities
with the Kaluza-Klein reduction, \cite{Mad89c}. In principle, 
this coupling can induce cancellation of divergences in the 
quantization: we plan to analyze in more details quantization
and renormalization of the proposed model in our future work.

\vskip0.7cm
{\bf Acknowledgment}\ \ 
This  work of was supported by ESF grants 2747 and 2749 through 
the Quantum Gravity Network.


\end{document}